\documentclass[superscriptaddress,groupedaddress,nofootnoteinbib,12pt]{article}
\usepackage{color}
\usepackage{graphicx}
\usepackage{dcolumn}
\usepackage{bm}
\usepackage{amssymb}
\usepackage{amsmath}
\usepackage{sectsty}
\usepackage{colortbl}
\usepackage{latexsym}
\usepackage{float}
\usepackage{ifthen}
\usepackage{caption,subfig}
\usepackage{enumerate}
\usepackage{url}
\usepackage{caption,subfig}	
\usepackage{jcappub}

\newcommand{\eff}{{\rm eff}}

\def\be{\begin{equation}}
\def\ee{\end{equation}}
\def\ba{\begin{eqnarray}}
\def\ea{\end{eqnarray}}
\def\beq{\begin{eqnarray}}
\def\eeq{\end{eqnarray}}

\def\d{\mathrm{d}}

\def\K{{\cal K}}

\def\L*{{\cal L}_*}
\def\L{\mathcal{L}}
\def\({\left(}
\def\){\right)}

\def\<{\langle}
\def\>{\rangle}

\def\cs2{c_{s}^{2}}

\def\be{\begin{equation}}
\def\ee{\end{equation}}
\def\ba{\begin{eqnarray}}
\def\ea{\end{eqnarray}}
\def\beq{\begin{eqnarray}}
\def\eeq{\end{eqnarray}}

\def\d{\mathrm{d}}

\def\K{{\cal K}}

\def\L*{{\cal L}_*}
\def\L{\mathcal{L}}
\def\({\left(}
\def\){\right)}

\def\<{\langle}
\def\>{\rangle}

\renewcommand{\thesubsection}{\arabic{section}.\arabic{subsection}}

\begin{document}
\begin{flushright}
YITP-15-XX \\
IPMU15-0004
\end{flushright}
\vspace{-1.03cm} 

\title{Cosmology in bimetric theory with an effective composite coupling to matter}

\author{A. Emir G\"umr\"uk\c c\"uo\u glu$^{a}$, Lavinia Heisenberg$^{b,c}$, Shinji Mukohyama$^{d,e}$ and Norihiro Tanahashi$^{f}$}
\affiliation{$^{a}$School of Mathematical Sciences, University of Nottingham, \\
University Park, Nottingham, NG7 2RD, UK}
\affiliation{$^{b}$Nordita, KTH Royal Institute of Technology and Stockholm University, \\
Roslagstullsbacken 23, 10691 Stockholm, Sweden}
\affiliation{$^{c}$Department of Physics \& The Oskar Klein Centre, \\
AlbaNova University Centre, 10691 Stockholm, Sweden}
\affiliation{$^{d}$Yukawa Institute for Theoretical Physics, Kyoto University, \\
Kyoto 606-8502, Japan}
\affiliation{$^{e}$Kavli Institute for the Physics and Mathematics of the Universe,\\
Todai Institutes for Advanced Study, University of Tokyo (WPI),\\
5-1-5 Kashiwanoha, Kashiwa, Chiba 277-8583, Japan}
\affiliation{$^{f}$Department of Applied Mathematics and Theoretical Physics, \\
University of Cambridge, Wilberforce Road, Cambridge CB3 OWA, UK}

\emailAdd{Emir.Gumrukcuoglu@nottingham.ac.uk} 
\emailAdd{Lavinia.Heisenberg@unige.ch}
\emailAdd{shinji.mukohyama@yukawa.kyoto-u.ac.jp}
\emailAdd{N.Tanahashi@damtp.cam.ac.uk}

\abstract{
We study the cosmology of bimetric theory with a composite matter coupling. We find two possible branches of background evolution. We investigate the question of stability of cosmological perturbations. For the tensor and vector perturbations, we derive conditions on the absence of ghost and gradient instabilities. For the scalar modes, we obtain conditions for avoiding ghost degrees. In the first branch, we find that one of the scalar modes becomes a ghost at the late stages of the evolution. Conversely, this problem can be avoided in the second branch. However, we also find that the constraint for the second branch prevents the doubly coupled matter fields from being the standard ingredients of cosmology. We thus conclude that a realistic and stable cosmological model requires additional minimally coupled matter fields.}

\maketitle

\section{Introduction}

Over the last century cosmology has progressively become an empirical scientific discipline. With the high precision achieved by the cosmological observations, cosmology is now considered as an indispensable arena to test fundamental physics. Nobel prizes have been awarded to the astonishing discoveries by the measurements of supernovae and Cosmic Microwave Background (CMB) radiation, which consolidated the standard model of cosmology. Another irreplaceable cosmological information is provided by galaxy surveys which is used in CMB secondary anisotropies and weak gravitational lensing.  ESA's PLANCK satellite releases very high precision measurements of the CMB and EUCLID mission will provide overwhelming amounts of data with very good control of systematical errors \cite{Ade:2013zuv,Weinberg:2012es,Amendola:2012ys}. All these probes allow us to disentangle the composition, structure and evolution of the Universe. Even if the standard model of cosmology accounts for most of the cosmological observations, there exist anomalies in tension to this model \cite{Ade:2013nlj} and theoretical discontent about the unnatural smallness of the observed cosmological constant \cite{Weinberg:1988cp}.\\

The aforementioned difficulties have initiated a major motivation to study modifications of gravity in the IR with additional dynamical fields. One of the most natural ways to modify gravity in the IR is to give the graviton a mass. In this context the de Rham--Gabadadze--Tolley (dRGT) theory has received much attention, being the first formulation of a non-linear covariant theory of massive gravity with the correct degrees of freedom \cite{deRham:2010ik,deRham:2010kj}. The potential is constructed in a way that guarantees the absence of the extra degree of freedom, dubbed the Boulware--Deser (BD) ghost \cite{Boulware:1973my}. The theory can be promoted to a massive bimetric theory by the inclusion of the kinetic term for the reference metric \cite{Hassan:2011zd}. After constructing consistent self-interacting spin-2 theories as in bigravity, a mandatory question to address is how these fields couple to the matter fields  \cite{Khosravi:2011zi, Akrami:2012vf,Akrami:2013ffa,Tamanini:2013xia,Akrami:2014lja, deRham:2014naa,deRham:2014fha,Yamashita:2014fga,Noller:2014sta,Schmidt-May:2014xla,Enander:2014xga,Solomon:2014iwa,Soloviev:2014eea,Heisenberg:2014rka,Mukohyama:2014rca}. In bi-gravity, both metrics are put on an equal footing, and coupling matter to both metrics simultaneously might thus appear natural at first sight. However, as shown in  \cite{deRham:2014naa} minimally coupling the matter sector to more than one metric generically introduces the propagation of ghost-like degrees of freedom already at the classical level. Moreover, quantum corrections at one loop destroy the special structure of the potential at an unacceptable low scale. A consistent way of coupling the matter sector to the two metric is to minimally couple the matter fields to just one metric, but not to both simultaneously. If one insists on coupling the matter sector to both metrics, a new composite effective metric constructed out of both metric was proposed in \cite{deRham:2014naa}. A nice feature of the coupling is that quantum corrections at one loop maintain the nice potential structure. Furthermore, this new matter coupling evades the no-go result for the flat Friedmann--Lema\^itre--Robertson--Walker (FLRW) background present in the original formulation of massive gravity and also allows the propagation of the five physical degrees of freedom of the graviton sector without ghost and gradient instabilities \cite{Gumrukcuoglu:2014xba}. From the lessons learned about the quantum corrections coming from matter loops, one can construct other classes of new effective metrics through which the matter field could couple to both metrics at the same time \cite{Heisenberg:2014rka}.\\

The absence of the BD ghost strongly relies on the relative tuning of the potential interactions. It is an unavoidable question to address whether or not this relative tuning is radiatively stable. These questions were addressed in \cite{deRham:2012ew,deRham:2013qqa} for massive gravity. In the decoupling limit the theory is protected from quantum corrections thanks to the non-renormalization theorem \cite{deRham:2012ew} (in the same way as the Galileon interactions are protected from quantum corrections due to a similar non-renormalization theorem \cite{Luty:2003vm,Nicolis:2004qq,Hinterbichler:2010xn,deRham:2012ew,Heisenberg:2014raa}). The antisymmetric structure of the potential interactions in the decoupling limit guarantees that any vertex will contribute at least with two additional momenta applied on the external leg to the transition amplitude and therefore the classical interactions remain untouched. Beyond the decoupling limit, the graviton loops do detune the specific structure of the potential interactions, however the mass of the introduced BD ghost is never below the cut-off scale of the theory and hence is harmless \cite{deRham:2013qqa}. The same detuning happens in the bimetric generalization of the theory \cite{Heisenberg:2014rka}.\\

The phenomenological aspects of the dRGT massive gravity have been widely studied, specially the potential impact on cosmology  \cite{deRham:2010tw,PhysRevD.84.124046,PhysRevLett.109.171101,deRham:2011by,Chamseddine:2011bu,Koyama:2011xz,Koyama:2011wx,Gumrukcuoglu:2011zh,Gratia:2012wt,Vakili:2012tm,Kobayashi:2012fz,Fasiello:2012rw,Volkov:2012zb,Tasinato:2012ze,Wyman:2012iw,Gratia:2013gka,DeFelice:2013bxa,Fasiello:2013woa,Heisenberg:2014kea,Comelli:2013tja,Motloch:2014nwa}. In the decoupling limit of the theory, it has been shown that the theory admits self-accelerating solutions \cite{deRham:2010tw}. What essentially happens is that the helicity-0 degree of freedom of the massive graviton forms a condensate whose energy density sources self-acceleration. Furthermore, small fluctuations around these self-accelerating backgrounds are stable. Another crucial result is that the fluctuations of the helicity-0 field do not couple to the fluctuations of the helicity-2 field (hence to the matter fields since it is the fluctuations of the helicity-2 field which couple to the matter field). Thus the cosmological evolution is exactly as in the standard $\Lambda$CDM model without the need of the Vainshtein mechanism \cite{deRham:2010tw}. Nevertheless, these self-accelerating solutions found in the decoupling limit do not guarantee the existence of full solutions with identical properties in the full theory. One can consider the solutions in the decoupling limit just as a transient state of the full solution. 

For the full theory, the construction of a realistic and stable cosmology has been a challenge so far. For dRGT massive gravity with Minkowski reference metric, there are no flat FLRW solution \cite{PhysRevD.84.124046} although self-accelerating open FLRW solutions exist \cite{Gumrukcuoglu:2011ew}. However, these self-accelerating solutions (and similar ones with other reference metrics) are known to have three instantaneous modes \cite{Gumrukcuoglu:2011zh} and lead to a nonlinear ghost instability \cite{PhysRevLett.109.171101}. Although for a de Sitter reference metric, other solutions have also been found \cite{Fasiello:2012rw,Langlois:2012hk}, these suffer from a Higuchi \cite{Higuchi:1986py} type ghost at high energies \cite{Fasiello:2012rw}. The attempts to circumvent these issues roughly fall into two categories: i) breaking the FLRW symmetries in the fiducial metric which do not affect the FLRW symmetries of the physical metric \cite{Gumrukcuoglu:2012aa,Koyama:2011xz,Koyama:2011yg,Chamseddine:2011bu,Gratia:2012wt,Kobayashi:2012fz,Volkov:2012cf}, or breaking these symmetries completely while relying on a cosmological screening mechanism \cite{PhysRevD.84.124046}; ii) extending the theory by adding new degrees of freedom \cite{Huang:2012pe,D'Amico:2012zv,DeFelice:2013tsa}.

The bimetric theory with dRGT tuning proposed by Ref.~\cite{Hassan:2011zd} can be seen to fall into the second category above. The cosmology of the bigravity theory has been already studied in a multitude of very interesting works \cite{Volkov:2011an,Volkov:2012cf,vonStrauss:2011mq,Akrami:2012vf,Akrami:2013pna,Maeda:2013bha,Volkov:2013roa,Akrami:2013ffa,Konnig:2013gxa,Aoki:2013joa,Comelli:2014bqa,Konnig:2014xva,Enander:2014xga,Lagos:2014lca,Enander:2015vja}. However, just like in massive gravity, attaining a stable cosmology is highly challenging. For minimally coupled matter fields and a small interaction term $m\sim H_0$, the theory admits several branches of solutions: a self-accelerating branch which has three instantenous modes and is thus unstable \cite{Comelli:2012db}, a branch where there is an early time gradient instability \cite{Comelli:2012db} and a branch where there is a crossing of a curvature singularity \cite{Comelli:2011zm} (dubbed ``infinite''branch in Refs.~\cite{Konnig:2014xva,Enander:2015vja}). In the latter branch, even if one neglects the consequences of the singularity, three degrees (two tensor and one scalar) become ghosts at the early stages of the evolution \cite{DeFelice:2014nja,Cusin:2014psa}. On the other hand, for a strongly interacting bimetric theory $m\gg H_0$, a viable but finely tuned solution exists \cite{DeFelice:2013nba,DeFelice:2014nja}.

In this work we will study the cosmological implications of the new effective metric proposed in \cite{deRham:2014naa} in the context of massive bigravity. The background evolution for one branch of solutions was already studied in \cite{Enander:2014xga} and we will push further this analysis to the level of the perturbations, while also paying attention to the other existing branches of solutions. After quickly reviewing the bimetric generalization of dRGT theory and clarifying our conventions and notations in Section \ref{sec:bigravity}, we first study the background equations of motion on FLRW space-time in Section \ref{sec:background_evolution}. We comment on the existence of two branches of solutions and their implications focusing on their early and late-time behaviors. In the same way as in massive gravity \cite{Gumrukcuoglu:2014xba}, the new coupling enables us to avoid the no-go theorem for FLRW solutions. In Section \ref{sec:perturbations} we turn our attention to the stability of the perturbations and show that even if the tensor perturbations can be maintained free of ghost and gradient instabilities, the vector and scalar perturbations unavoidably yield gradient and ghost instabilities at early and late time evolutions respectively in one of the branches of solutions. We then conclude in Section \ref{sec:conclusion}.\\

Throughout the paper, we will denote traces by $[...]$, for example the contractions of rank-2 tensors as ${\cal K}^{\mu}_{~\mu}=[{\cal K}]$,~ ${\cal K}^{\mu}_{~\nu}{\cal K}^{\nu}_{~\mu}=[{\cal K}^2]=({\cal K}_{\mu\nu})^2$,~ ${\cal K}^{\mu}_{~\nu}{\cal K}^{\nu}_{~\rho}{\cal K}^{\rho}_{~\mu}=[{\cal K}^3]=({\cal K}_{\mu\nu})^3$~etc.


\section{Massive bimetric gravity}
\label{sec:bigravity}
We will start this section with a review of the massive bimetric theory and collect all the important quantities. 
 Our action is given by
\begin{eqnarray}\label{action_BG_effcoupl}
\mathcal{S} &=& \int \mathrm{d}^4x \left[ \frac{M_g^2}{2} \sqrt{-g}\left(R[g]+2m^2\sum_n \alpha_n{\cal U}[\cal K]  \right)+\sqrt{-g_{\rm eff}}\mathcal{L}_\chi(g_{\rm eff},\chi)\right]\,, \nonumber\\ 
&+&  \int \mathrm{d}^4x \left[ \frac{M_f^2}{2} \sqrt{-f}R[f] +\sqrt{-f}\mathcal{L}_{\rm matter}[f]  +  \sqrt{-g}\mathcal{L}_{\rm matter}[g]  \right]
\end{eqnarray}
with the very specific potential interactions \cite{deRham:2010ik,deRham:2010kj}
\begin{eqnarray}
\mathcal{U}_0[\mathcal{K}] &=&\frac{1}{24} \mathcal{E}^{\mu\nu\rho\sigma}  \mathcal{E}_{\mu\nu\rho\sigma} =1\nonumber\\
\mathcal{U}_1[\mathcal{K}] &=&\frac{1}{6} \mathcal{E}^{\mu\nu\rho\sigma}  \mathcal{E}^{\alpha}_{\;\;\;\nu\rho\sigma} \K_{\mu\alpha} =[\K] \nonumber\\
\mathcal{U}_2[\mathcal{K}] &=&\frac{1}{4} \mathcal{E}^{\mu\nu\rho\sigma}  \mathcal{E}^{\alpha\beta}_{\;\;\;\;\; \rho\sigma} \K_{\mu\alpha} \K_{\nu\beta} = \frac{1}{2}\left( [\K]^2-[\K^2]\right), \nonumber\\
\mathcal{U}_3[\mathcal{K}] &=& \frac{1}{6} \mathcal{E}^{\mu\nu\rho\sigma}  \mathcal{E}^{\alpha\beta\kappa}_{\;\;\;\;\;\;\; \sigma} \K_{\mu\alpha} \K_{\nu\beta}  \K_{\rho\kappa}=\frac{1}{6}\left([\K]^3-3[\K][\K^2]+2[\K^3]\right),  \nonumber\\
\mathcal{U}_4[\mathcal{K}] &=& \frac{1}{24} \mathcal{E}^{\mu\nu\rho\sigma}  \mathcal{E}^{\alpha\beta\kappa\gamma} \K_{\mu\alpha} \K_{\nu\beta}  \K_{\rho\kappa}  \K_{\sigma\gamma} =\frac{1}{24} \left([\K]^4-6[\K]^2[\K^2]+3[\K^2]^2+8[\K][\K^3]-6[\K^4]\right) \nonumber\\
\end{eqnarray}
where the square brackets denote trace operation and $\mathcal{E}$ is the Levi-Civita tensor. The matrix $\K$ is given by
\begin{equation}
\K^\mu _\nu[g,f] =\delta^\mu_\nu - \left(\sqrt{g^{-1}f}\right)^\mu_\nu \,.
\end{equation}
In \cite{deRham:2014naa} a new effective composite coupling was proposed
\begin{equation}
g^\eff_{\mu\nu} \equiv \alpha^2 g_{\mu\nu}+2\,\alpha\,\beta\, g_{\alpha\mu} \left(\sqrt{g^{-1}f}\right)^\alpha_{\nu} + \beta^2 f_{\mu\nu}\,.
\label{eq:geff}
\end{equation}
In this paper we will consider a scalar field $\chi$ with a generic kinetic term of the form 
\begin{equation}
\mathcal{L}_{\chi} =\sqrt{-g_\text{eff}}\,P(X_\chi)\,,
\label{eq:chiaction}
\end{equation}
where $X_\chi$ stands for
\begin{eqnarray}
  X_\chi \equiv -g_\text{eff}^{\mu\nu}\partial_{\mu}\chi\partial_{\nu}\chi
\,.
\label{matter}
\end{eqnarray}
The energy density, the pressure and the sound speed of the scalar field correspond to
\begin{equation}
 \rho_\chi \equiv 2P(X_\chi)'X_\chi - P(X_\chi), \qquad  P_\chi \equiv P(X_\chi), \qquad
 c_\chi^2 \equiv \frac{P(X_\chi)'}{2P(X_\chi)''X_\chi+P(X_\chi)'}\,,
\end{equation}  
where prime denotes derivative with respect to the argument. Instead of a k-essence field \cite{ArmendarizPicon:2000ah} as in (\ref{eq:chiaction}), one could also consider a more general scalar field, for instance a Galileon field \cite{Nicolis:2008in} or a Horndeski field \cite{Horndeski:1974wa} or even a general vector field \cite{Jimenez:2013qsa,Heisenberg:2014rta,Tasinato:2014eka}, but just for simplicity we will restrict ourselves to a k-essence field. In our setup we will consider two cosmological constants as a place-holder for the matter fields which either couple only to $g$ or to $f$%
\footnote{
At this stage, we would like to point out that the introduced cosmological constants $\Lambda_g$ and $\Lambda_f$ serve only as a technical tool to keep track of the energy densities of additional fields and in no way should be thought as a real representation of physical matter fields. 
It would be important to study cosmological solutions and perturbations on them 
taking more realistic matter fields into account.
}
\begin{equation}
\mathcal{L}^g_{\rm matter}=-M_g^2\Lambda_g\,,  \;\;\;\;\;\;\;\;\;\;\;\;\;\;\;\;     \mathcal{L}^f_{\rm matter}=-M_f^2\Lambda_f\,.
\end{equation}
In the following section we will first study the background equations and discuss the different branches of solutions together with their early and late time regime behaviors before moving on to the perturbations.


\section{Background evolution}\label{sec:background_evolution}

The bimetric theory in the original formulation without the composite coupling has been already intensively studied in the literature.
In the context of the new effective coupling, Enander, Solomon, Akrami and M\"ortsell have found cosmological solutions that resembles $\Lambda$CDM for a given choice 
of parameters \cite{Enander:2014xga}. This opens up new and exciting avenues to explore. In this work, we will perform perturbations about these cosmological solutions and discuss their stability. We assume homogeneous and isotropic flat FLRW metrics
\begin{eqnarray}
ds_g^2&=&-n_g^2 dt^2 +a_g^2 \delta_{ij} dx^idx^j\,,  \nonumber\\
ds_f^2&=&-n_f^2 dt^2 + a_f^2  \delta_{ij} dx^idx^j\,.
\end{eqnarray}
Whit our Ansatz the effective metric becomes
\begin{equation}
ds^2_\eff = -N^2_\eff dt^2+a_\eff^2 \delta_{ij}dx^idx^j\,,
\end{equation}
where the effective lapse $N_{\eff}$ and scale factor $a_{\eff}$ are defined as
\begin{equation}
N_\eff \equiv \alpha\,n_g+\beta\,n_f\,,\qquad
a_\eff \equiv \alpha\,a_g+\beta\,a_f\,.
\end{equation}
Compatible with our above homogeneous and isotropic Ansatz the $\chi$ field also only depends on time $\chi=\chi(t)$. The introduction of the three Hubble rates
\begin{equation}\label{shortcuts}
H_g\equiv\dot{a}_g/(n_ga_g), 
\qquad 
H_f\equiv \dot{a}_f/(n_fa_f),
\qquad
 H_\text{eff} \equiv \frac{\dot a_\text{eff}}{N_\text{eff}\, a_\text{eff}}
\end{equation}
and the ratios of the background scale factors and lapse functions 
\begin{equation}
 A  \equiv  \frac{a_f}{a_g}\,, 
\qquad 
r \equiv \frac{n_f a_g}{n_g a_f}\,, 
\end{equation}
and the following function encoding the potential interactions
\begin{equation}
 U(A)\equiv -\alpha_0+4(A-1)\alpha_1-6(A-1)^2\alpha_2+4(A-1)^3\alpha_3-
(A-1)^4\alpha_4
\end{equation}
will be beneficial for the exposure of the equations. 
Throughout this paper, we assume $A>0$ and $r>0$.
The action (\ref{action_BG_effcoupl}) in the mini-superspace becomes:
\begin{eqnarray}
\frac{\mathcal{S}}{V} &=& M_g^2\int dt \,a_g^3 n_g\,\Bigg\{-\Lambda_g-3H_g^2-m^2\rho_{m,g}\Bigg\} + M_f^2\int dt \,a_f^3 n_f\,\Bigg\{-\Lambda_f-3H_f^2-m^2\frac{M_g^2}{M_f^2}\rho_{m,f}\Bigg\}\nonumber\\
&&+\int dt \,a_\eff^3 N_\eff P\left( \frac{\dot\chi^2}{N_\eff^2} \right)\,,
\label{eq:minisuperspace}
\end{eqnarray}
where for convenience we used the following functions as effective energy densities from the mass term instead of the parameters $\alpha_n$ 
\begin{equation}
\rho_{m,g} (A)\equiv U(A)-\frac{A}{4}\, U_{,A} \,,\qquad
\rho_{m,f} (A)\equiv \frac{1}{4A^3}\, U_{,A}\,.
\end{equation}
We can now compute the background equations by varying the action (\ref{eq:minisuperspace}) with respect to $n_g$, $n_f$, $a_g$, $a_f$ and $\chi$. The variation with respect to the two
lapses yield the Friedmann equations
\begin{eqnarray}
3\,H_g^2 &=& \Lambda_g + m^2 \rho_{m,g} +\frac{\alpha\,a_\eff^3}{M_g^2\,a_g^3}\rho_\chi\,, \nonumber\\
3\,H_f^2 &=& \Lambda_f + m^2\frac{M_g^2}{M_f^2} \rho_{m,f} +\frac{\beta\,a_\eff^3}{M_f^2\,a_g^3\,A^3}\rho_\chi\,, 
\label{eq:eqN}
\end{eqnarray}
On the other hand, varying the mini-superspace action (\ref{eq:minisuperspace}) with respect to the scale factors $a_g$ and $a_f$, then using the Friedmann equations results in the following acceleration equations
\begin{eqnarray}
\frac{2\,\dot{H_g}}{n_g}&=& m^2\,J\,A\,(r-1) - \frac{\alpha\,a_\eff^3}{M_g^2a_g^3}\left[\rho_\chi+ \frac{N_\eff/a_\eff}{n_g/a_g}P_\chi\right]\,, \nonumber\\
\frac{2\,\dot{H_f}}{n_f}&=& -\frac{m^2}{A^3}\frac{M_g^2}{M_f^2}\,\frac{J}{r}\,(r-1) - \frac{\beta\,a_\eff^3}{M_f^2a_g^3A^3}\left[\rho_\chi+\frac{1}{r} \frac{N_\eff/a_\eff}{n_g/a_g}P_\chi\right]\,,
\label{eq:eqa}
\end{eqnarray}
where we have defined the function $J=\frac{1}{3}\partial_A \rho_{m,g}$ for clarity. Last but not least, the equation of motion for the $\chi$ field is just the standard conservation equation for a field minimally coupled to the $g_\eff$ metric
\begin{equation}
\left[\frac{1}{c_\chi^2}\frac{1}{N_\eff}\,\frac{\partial}{\partial t}\,\left(\frac{\dot{\chi}}{N_\eff}\right) +3H_\eff\,\frac{\dot{\chi}}{N_\eff}\right](\rho_\chi+P_\chi)=0\,.
\label{eq:eqchi}
\end{equation}
Combining the equations for the lapse $n_g$, the acceleration equation for $a_g$ and the equation of motion for the scalar field $\chi$ yields the following constraint equation
\begin{equation}
\left(m^2  J -  \frac{\alpha\beta a_\eff^2}{M_g^2a_g^2} P_\chi \right)(H_g-A H_f) = 0\,.
\label{constraint}
\end{equation}
From this constraint we immediately see that there are two branches of solutions.
\subsection{Branch I}
The first branch of solutions corresponds to the case in which 
\begin{equation}
H_g-A H_f=0 
\end{equation}
in the constraint equation (\ref{constraint}). Combining the equations of motion for the lapses in (\ref{eq:eqN}) gives rise to the algebraic relation
\begin{equation} 
m^2 \left(\rho_{m,g}-\frac{A^2M_g^2}{M_f^2}\,\rho_{m,f}\right) +\Lambda_g - \Lambda_f A^2=- \frac{\rho_\chi}{M_g^2} \left( \alpha - \frac{\beta M_g^2}{M_f^2 A} \right)
\left( \frac{a_\text{eff}}{a_g} \right)^3
\,.
\label{Fconst1}
\end{equation}
Another additional constraint comes from $\frac{\partial}{\partial t}\left(H_g-A H_f\right)=0$. Using the acceleration equations (\ref{eq:eqa}), this constraint can be equally written as
\begin{equation}
2 (r -1) W=\frac{(a_\eff/a_g)^3}{M_g^2}\left[\left( \alpha - \frac{\beta M_g^2\,r}{M_f^2 A} \right)\rho_\chi +\left( \alpha - \frac{\beta M_g^2}{M_f^2 A} \right)\frac{N_\eff/a_\eff}{n_g/a_g}P_\chi\right]
\,, 
\label{Fconst2}
\end{equation}
where we have defined 
\begin{equation}
W\equiv \frac{m^2(M_g^2 +M_f^2 A^2)J}{2M_f^2 A}-H_g^2
\,.
\end{equation}
These two consistency relations ensure that the $f_{\mu\nu}$ metric evolves according to the constraint $H_f = H/A$, by fixing $A$ and $r$. 

The relation (\ref{Fconst1}) can be used to analyze the early and late time limits of the cosmological evolution. Assuming no bare cosmological constant ($\Lambda_g=\Lambda_f=0$) and that at late times, the contribution from the two-metric interaction term $\propto m^2$ dominates the expansion, we require
\begin{equation}
\frac{\alpha\,a_\eff^3}{a_g^3\,m^2M_g^2}\,\rho_\chi \ll1\,,\qquad {\rm late~times}\,.
\end{equation}
This requirement can be consistent, provided that the energy density of the matter field $\chi$ decreases with expansion. For instance, for pressureless dust, $\rho_\chi \propto a_\eff^{-3}$. If the late time acceleration purely stems from the $m^2\rho_{m,g}$ term, i.e.\ parametrically $m^2\sim H_0^2$, this means that the contribution from $\rho_\chi$ is already negligible at the present time. In this late time regime, the consistency relation (\ref{Fconst1}) then gives
\begin{equation}
m^2 \left(\rho_{m,g}-\frac{A^2M_g^2}{M_f^2}\,\rho_{m,f}\right) \simeq 0\,,
\label{branchIeq}
\end{equation}
or, at late times, $A$ approaches to a constant value determined by the root of the above equation. 
Then equation (\ref{Fconst2}) implies that $r\simeq 1$.
This asymptotic regime is self-accelerating.

Similarly, at early times, one has
\begin{equation}
\frac{\alpha\,a_\eff^3}{a_g^3\,m^2M_g^2}\,\rho_\chi \gg1\,,\qquad {\rm early~times}\,.
\end{equation}
If for the sake of argument, we assume no other matter field minimally coupled to a single metric, 
the consistency relation~(\ref{Fconst1}) implies that $\alpha - \frac{\beta M_g^2}{M_f^2 A}\simeq 0$ \cite{Enander:2014xga}.

Of course, these constraints will be drastically modified if there is any ``regular'' matter sector, in the sense that it couples minimally to only one of the two metrics. In this case, the right-hand side of (\ref{Fconst1}) may be dominated by these matter fields rather than the mixed coupled one, reducing to \cite{DeFelice:2014nja}.

\subsection{Branch II}
The second branch of solutions corresponds to the case where 
\begin{equation}
m^2  J -  \frac{\alpha\beta a_\eff^2}{M_g^2a_g^2} P_\chi =0
\label{eq:branch2}
\end{equation}
coming from the constraint equation (\ref{constraint}). The dynamics of this branch is drastically different than the one for usual minimal matter couplings ($\alpha=0$ or $\beta=0$), in the sense that the matter sector directly enters the constraint. On the other hand, for an arbitrary matter field coupling and non-zero pressure, the solutions $A$ are generically time dependent. In this branch, the acceleration equations (\ref{eq:eqa}) yield an interesting consistency relation 
\begin{equation}
\frac{\dot{H}_f}{\dot{H}_g} = \frac{M_g^2\beta\,r}{M_f^2\,\alpha\,A^2}\,.
\label{constII_1}
\end{equation}
Also, requiring that the constraint (\ref{eq:branch2}) is satisfied throughout the evolution, i.e. 
$\frac{\partial}{\partial t} \bigl(m^2  J -  \alpha\beta (a_\eff/(M_ga_g))^2 P_\chi\bigr)=0$
gives another constraint equation in this branch
\begin{equation}
\frac{2m^2 a_g^2 n_g}{(r-1)A}
\left[
-rAJ\left( H_g- A H_f \right) + \Gamma \left(H_g - r A H_f\right)
\right]
+ \frac{\alpha \beta}{M_g^2} a_\text{eff}^2 H_\text{eff}N_\text{eff}\left[
2P_\chi - 3 c_\chi^2 (\rho_\chi + P_\chi)
\right]=0\,,
\label{constII_2}
\end{equation}
where 
\begin{equation}
   \Gamma \equiv A J + \frac{(r -1) A^2}{2} J_{,A}
  \,.
\label{Gamma} 
\end{equation}

Let us briefly discuss the early and late time limits in this branch.
Similarly to the arguments for the branch-I solutions we made above, we assume $\Lambda_g = \Lambda_f = 0$.
Then, at late times we assume that $\rho_\chi$ and $P_\chi$ decays to values sufficiently smaller than $m^2M_g^2$:%
\footnote{
We remark that this assumption is not the unique one that should hold in the late time,
and there could be another late time regime in which these conditions are not satisfied.
In fact, when parameters $\alpha_i$ are such that $J(A)>0$ for any $A$, there is a minimum value of $P_\chi$ which can parametrically be of order $M_g^2 m^2$. In this case, there is no regime where the $\chi$ is neither dominant nor sub-dominant. As we expand in the next Section, these backgrounds can evade the strong-coupling problem we will encounter.}
\begin{equation}
\frac{\alpha\beta a_\text{eff}^2}{m^2M_g^2 a_g^2 }P_\chi 
\ll 1
\, , \qquad
\frac{\alpha\beta a_\text{eff}^2}{m^2M_g^2 a_g^2 }\rho_\chi
 \ll 1~.
\label{lowElimit_II}
\end{equation}
In this regime, equation (\ref{eq:branch2}) implies $J\ll 1$, and then $A$ converges to the constant values defined $J=\frac{1}{3}\partial_A \rho_{m,g}=0$, that is,
\begin{equation}
A=\frac{\alpha_2+2\alpha_3+\alpha_4\pm\sqrt{\alpha_2^2+\alpha_3(\alpha_3-\alpha_1+\alpha_2)-(\alpha_1+\alpha_2)\alpha_4}}{\alpha_3+\alpha_4}\,.
\label{eq:Asol-dRGT}
\end{equation}
As a result, the effective energy densities $\rho_{m,g}$ and $\rho_{m,f}$ are also constant, leading to a late time self-acceleration terms in the Friedmann equations for both metrics (\ref{eq:eqN}). Then, if we further assume that $A,\,r = {\cal O}(1)$, the constraint (\ref{constII_2}) reduces to
\begin{equation}
\frac{\Gamma}{(r-1)}\left(H_g - r A H_f\right) 
\simeq
\frac12
A^2J_{,A}
\left(H_g - r A H_f\right) 
\ll 1~,
\end{equation}
which suggests that $H_g \simeq r A H_f$ in the late time regime in a generic case where $J_{,A}={\cal O}(1)$. This can be also easily seen by the relation 
\begin{equation}\label{relationdotA}
\frac{\dot A}{n_g A}=H_f r A - H_g
\end{equation}
which for a constant $A$ yields the said relation at late times. 

In dRGT massive gravity and bimetric theory, this behavior with constant $A$ (\ref{eq:Asol-dRGT}) is valid at any stage of the evolution since the constraint is unaffected by the minimally coupled matter fields. Although in this case, this branch is known to contain three degrees of freedom with vanishing kinetic terms \cite{Gumrukcuoglu:2011zh,Comelli:2012db,Comelli:2014bqa}, we will show in the following Section that for a composite matter coupling, this problem can be evaded even at the late stages of the evolution where the situation is similar to standard dRGT.

The early time regime could be defined by the condition that the $\chi$ field is the dominant component of the evolution, i.e.,
\begin{equation}
\frac{\alpha\beta a_\text{eff}^2}{m^2M_g^2 a_g^2 }P_\chi 
\gg 1
\, , \qquad
\frac{\alpha\beta a_\text{eff}^2}{m^2M_g^2 a_g^2 }\rho_\chi
 \gg 1~.
\label{largeElimit_II}
\end{equation}
To see if these conditions could be satisfied, it is useful to rewrite
 the constraint equation (\ref{eq:branch2}) into
\begin{equation}
P_\chi = \frac{M_g^2 m^2}{\alpha\beta}
\frac{\alpha_1 + 3\alpha_2+3\alpha_3 + \alpha_4 -2(\alpha_2+2\alpha_3 + \alpha_4)A +(\alpha_3 + \alpha_4)A^2}{(\alpha + \beta\,A)^2}~.
\label{eq:boundedP}
\end{equation}
We may view this equation as an algebraic equation to determine $A$ for a given $P_\chi$.
Since we assumed $A>0$, the right-hand side is bounded 
and thus solutions to this equation exist only when parametrically $|P_\chi|\lesssim m^2M_g^2$ 
 if $\alpha, \beta, \alpha_{0,1,2,3,4} = {\cal O}(1)$
.%
\footnote{
Here, we suppose that $\alpha\beta>0$, which is necessary for the
stability of the vector and scalar perturbations as we shall see in
(\ref{eq:noghostvector}) (with (\ref{eq:branch2}) imposed) and (\ref{eq:noghostscalar}).
}
During the course of the evolution, the pressure of the $\chi$ matter field reaches a maximum value. At this extremal value of the pressure, we find from Eq.~(\ref{constII_2}) that
\begin{equation}
\frac{\dot{A}}{n_g}\Big\vert_{P_\chi = P_\chi(max)} = -\frac{(\alpha+A\,\beta)H_g}{\beta}\,,
\end{equation}
which further implies $H_\text{eff}\Big\vert_{P_\chi = P_\chi(max)} =0$. In other words, the effective metric bounces at this instant. As a result of this bounce, the $\chi$ matter field never dominates the expansion. This result suggests that one cannot have a realistic cosmological model in this branch, unless ``regular'' matter fields coupling minimally to either of the two metrics exist. When such matter fields are included, the bound given by Eq.~(\ref{eq:boundedP}) will not change, but it will still be possible to have a matter or radiation dominated stages using the added fields.


\section{Perturbations}\label{sec:perturbations}

The previous section was dedicated to the study of the background equations of motion and the existence of flat FLRW solutions in the presence of the new coupling to the matter sector in the framework of bigravity. This new coupling had tremendous implications in the framework of massive gravity since the original formulation does not accommodate for flat FLRW solutions. It was shown that the new coupling  can circumvent this no-go result and that the BD ghost is not present around these flat FLRW solutions \cite{deRham:2014naa}. Additionally, the decoupling limit analysis had shown that the theory is free of the BD ghost at least up to the strong coupling scale $\Lambda_3^3=M_g m^2$ and hence provides a valid effective field theory up to this scale. Furthermore, the perturbations around these cosmological solutions were investigated in \cite{Gumrukcuoglu:2014xba} and the stability conditions  for all the propagating six degrees of freedom were explored. The theory also avoids the no-go result of vanishing kinetic terms for the physical degrees of freedom in the case of the standard coupling. In this work we will have a similar analysis as in  \cite{Gumrukcuoglu:2014xba} but for the case of bigravity. In order to determine the stability conditions around the flat FLRW solutions, we consider the following Ansatz for the two dynamical fields
\begin{eqnarray}\label{AnsatzPerturbations}
&& g_{\mu\nu}dx^{\mu}dx^{\nu} 
  = -n_g^2 (1+2\Phi)\,dt^2
  + 2n_g a_g V_i \,dt\,dx^i
  + a^2(\delta_{ij}+h_{ij}) dx^idx^j\,, \nonumber\\
&& f_{\mu\nu}dx^{\mu}dx^{\nu} 
  = -\ n_f^2 (1+2 \varphi)\, dt^2
  + 2 n_f a_f v_i\, dt\,dx^i
  + a_f^2(\delta_{ij}+\gamma_{ij}) dx^idx^j\,, 
\end{eqnarray}
where the perturbations ($\Phi$, $V_i$, $h_{ij}$) and ($\varphi$, $v_i$, $\gamma_{ij}$) are functions of time and space. We will further decompose the tensor perturbations into their trace and traceless parts
\begin{equation}
 h_{ij} = \frac{1}{3}\delta_{ij}h  + h^T_{ij}\,,\quad
 \gamma_{ij} = \frac{1}{3}\delta_{ij}\gamma+ \gamma^T_{ij}\,,
\end{equation}
Furthermore, we consider the following perturbation for the matter sector $\chi$ 
\begin{equation}
\chi=\chi_0(t)+ \delta\chi\,.
\end{equation}
It will be convenient to decompose the perturbations in Fourier modes with respect to the spatial coordinates 
\begin{equation}
\mathcal{Q}=\int \frac{\d^3k}{(2\pi)^{3/2}}\mathcal{Q}_{\vec{k}}(t) \exp(i\vec{k}\cdot\vec{x}) +c.c. \, ,
\end{equation}
where $\mathcal{Q}$ represents the perturbations ($\Phi$, $V_i$, $h_{ij}$) and ($\varphi$, $v_i$, $\gamma_{ij}$) respectively. We will perform the stability analysis of the perturbations for each sector separately and focus on the conditions arising from the absence of ghost and/or gradient instabilities. 

\subsection{Tensor perturbations}
First of all, we will start with the analysis of the transverse-traceless part of the metric fluctuations. Our Lagrangian (\ref{action_BG_effcoupl}) together with the Ansatz (\ref{AnsatzPerturbations}) for the metric perturbations in Fourier modes becomes 
\begin{multline}
\label{tensorPerturbations}
\mathcal{S}^{(2)}_{\rm tensor} 
=
\frac{M_g^2}{8} \int d^3kdt \,n_g a_g^3 \Bigg[
\frac{\dot{h}_{ij,\vec{k}}^\star \dot{h}^{ij}_{\vec{k}}}{n_g^2}-\frac{k^2}{a_g^2}h_{ij,\vec{k}}^\star h^{ij}_{\vec{k}} 
+
\frac{M_f^2\, r \, A^4}{M_g^2}
\left(\frac{\dot{\gamma}_{ij,\vec{k}}^\star \dot{\gamma}^{ij}_{\vec{k}}}{n_f^2} -\frac{k^2}{a_f^2}\gamma_{ij,\vec{k}}^\star \gamma^{ij}_{\vec{k}} 
\right) 
\\
- m^2_\text{eff} \left(h^{ij}_{\vec{k}}-\gamma^{ij}_{\vec{k}}\right)\left(h_{ij,\vec{k}}^\star -\gamma_{ij,\vec{k}}^\star \right)
\Bigg]\, ,
\end{multline}
where $m^2_\text{eff}$ stands for the shortcut notation
\begin{equation}
m^2_\text{eff} 
\equiv 
m^2\Gamma 
- \alpha \beta A \left(\frac{a_\text{eff}}{a_g}\right)^2 
\frac{N_\text{eff}/a_\text{eff}}{n_g/a_g}\frac{P_\chi}{M_g^2}
=
m^2\Gamma 
- \alpha \beta A 
(\alpha + \beta A)(\alpha + \beta r A)
\frac{P_\chi}{M_g^2}
~,
\label{meff}
\end{equation}
and 
$h_{ij}$ and $\gamma_{ij}$ are assumed to be transverse-traceless in this section.
As one can see in the expression of the quadratic action (\ref{tensorPerturbations}), the tensor perturbations have the right sign for the kinetic term. The tensor modes do not impose any further constraint, since
 $r>0$ and $A>0$ already. They also do not give rise to any gradient instabilities. 
In the late time regime, the quantity $\frac{1+M_f^2 A^2/M_g^2}{M_f^2 A^2 / M_g^2}m^2_\text{eff}$ can be identified as the effective graviton mass squared \cite{DeFelice:2014nja}.
We discuss its properties in the next section. Even though the tensor perturbations are free of ghost and gradient instabilities, there can be tachyonic instabilities when $m^2_\text{eff}<0$. However, the parameters of the model can be chosen such that the timescale of the instability would be harmless. For more details on the tachyonic instability please see Ref.~\cite{Comelli:2015pua}.

\subsection{Vector perturbations}
We now concentrate on the vector perturbations. We decompose the metric perturbations as
\begin{eqnarray}
&V_{i,\vec{k}} = B_{i,\vec{k}}\,,\qquad 
&h_{ij,\vec{k}}=\frac{i}{2}\left(k_i E_{j,\vec{k}}+ k_j  E_{i,\vec{k}}\right)\,,\nonumber\\
& v_{i,\vec{k}} = b_{i,\vec{k}}\,,\qquad
&\gamma_{ij,\vec{k}}=\frac{i}{2}\left(k_i S_{j,\vec{k}}+ k_j  S_{i,\vec{k}}\right)\,,
\end{eqnarray}
where $\partial^i E_i = \partial^i B_i= \partial^i b_i=\partial^i S_i=0$. After taking into account the background equations of motion, the quadratic action of the vector perturbations becomes
\begin{align}
&\mathcal{S}_{\rm vector}^{(2)}  = \frac{M_g^2}{8}\int d^3kdt\,n_g a_g^3
 \Bigg[\frac{k^2}{2}\left(\frac{\dot{E}_{\vec{k}}^i}{n_g}
-\frac{2B_{\vec{k}}^i}{a_g}\right)\left(\frac{\dot{E}^\star_{i,\vec{k}}}{n_g}-\frac{2 B^\star_{i,\vec{k}}}{a_g}\right)
\notag \\
&
+\frac{k^2M_f^2A^4r}{2M_g^2}
\left(\frac{\dot{S}_{\vec{k}}^i}{n_f}-\frac{2 b^i_{\vec{k}}}{a_f}\right)
\left(\frac{\dot{S}^\star_{i,\vec{k}}}{n_f}-\frac{2b^\star_{i,\vec{k}}}{a_f}\right)
-
\frac{k^2m^2_\text{eff}}{2}
(E^i_{\vec{k}}-S^i_{\vec{k}})(E^\star_{i,\vec{k}}-S^\star_{i,\vec{k}})
\nonumber\\
&
+
\frac{2m^2_\text{eff}}{c_V^2}
(B^i_{\vec{k}}-r b^i_{\vec{k}})(B_{i,\vec{k}}^\star-r b^\star_{i,\vec{k}})
\Bigg]\,,
\label{actvec}
\end{align}
where we have introduced the following definition for convenience 
\begin{align}
c_V^2 &\equiv
\frac{r+1}{2}\frac{m^2_\text{eff}}{A}
\left[
m^2J 
 + \alpha\beta\left(\frac{a_\text{eff}}{a_g}\right)^2
\left(
\frac{1}{r+1}\frac{N_\text{eff}/a_\text{eff}}{n_g/a_g}\frac{\rho_\chi+P_\chi}{M_g^2}
+\frac{\rho_\chi}{M_g^2}
\right)
\right]^{-1}~.
\label{cV}
\end{align}
This quantity is proportional to the squared sound speed as discussed below.
Na\"ively counted there are four vector modes ($B_i$, $b_i$, $S_i$ and $E_i$). However, the vector modes $B_i$ and $b_i$ are not dynamical degrees of freedom and thus we can use their equations of motion in order to integrate them out. By doing so we obtain for the non-dynamical degrees of freedom
\begin{equation}
B_{i,\vec{k}} = a_g\left[\frac{\dot{E}_{i,\vec{k}}}{2 n_g}
-\frac{{\mathfrak C}}{2n_g}(\dot{E}_{i,\vec{k}}-\dot{S}_{i,\vec{k}})\right]\,,\qquad
b_{i,\vec{k}} = a_g\left[\frac{\dot{S}_{i,\vec{k}}}{2 n_gr}+\frac{M_g^2{\mathfrak C}}{2M_f^2 n_g A^2}(\dot{E}_{i,\vec{k}}-\dot{S}_{i,\vec{k}})\right]\,,
\label{Bbsol}
\end{equation}
where ${\mathfrak C}$ is given by
\begin{equation}
{\mathfrak C} \equiv 
\left[
\frac{k^2}{a_g^2}\frac{c_V^2}{m_\text{eff}^2} 
+ 
\frac{M_g^2}{M_f^2A^2}r+1 
\right]^{-1}
\,,
\label{calC}
\end{equation}
Integrating out the non-dynamical vector modes $B_i$ and $b_i$ through equations (\ref{Bbsol}) results in
\begin{align}
\mathcal{S}_{\rm vector}^{(2)} 
=
\frac{M_g^2}{8}\int d^3k\,dt\,n_g a_g^3
\left(
{\mathfrak C} \frac{\dot{\cal E}^i_{\vec{k}}\dot{\cal E}_{i,\vec{k}}^\star }{n_g^2}
- \frac{M_f^2 A^2 /M_g^2}{1+M_f^2A^2/M_g^2} m^2_\text{eff}
 {\cal E}^i_{\vec{k}}{\cal E}_{i,\vec{k}}^\star
\right)
\notag \,.
\end{align}
where we defined the gauge invariant combination 
\begin{equation}
 {\cal E}_{i,\vec{k}} \equiv 
   (E_{i,\vec{k}} - S_{i,\vec{k}})k
\,.
\end{equation}
For the absence of ghost instabilities we have to impose that the kinetic term has the right sign. As one can see from the expression for the quadratic action of the vector perturbations,
this is guaranteed by the requirement ${\mathfrak C}>0$.
A sufficient condition for this to hold is 
\begin{equation}
\frac{c_V^2}{m_\text{eff}^2}
= 
\frac{r+1}{2A}
\left[
m^2J 
 + \alpha\beta\left(\frac{a_\text{eff}}{a_g}\right)^2
\left(
\frac{1}{r+1}\frac{N_\text{eff}/a_\text{eff}}{n_g/a_g}\frac{\rho_\chi+P_\chi}{M_g^2}
+\frac{\rho_\chi}{M_g^2}
\right)
\right]^{-1}
>0~,
\label{eq:noghostvector}
\end{equation}
which is satisfied if $J>0$, $\rho_\chi>0$ and $\rho_\chi + P_\chi >0$.
Also the squared sound speed for the vector mode is given by
$\frac{M_f^2 A^2 /M_g^2}{1+M_f^2A^2/M_g^2} c_V^2$ and it should be positive to avoid the gradient instability.
For this stability, on top of the conditions to avoid the ghost instability, another condition $m_\text{eff}^2>0$ needs to be satisfied.

Let us briefly study the late time limit discussed in section~\ref{sec:background_evolution}.
In this limit, $m_\text{eff}^2$ behaves for both solution branches as
\begin{equation}
m_\text{eff}^2 \simeq m^2 \Gamma~,
\end{equation}
and $c_V^2$ behaves as 
\begin{equation}
c_V^2 =
\begin{cases}
\frac{\Gamma}{A\, J} 
& (\text{Branch I})
\\
\frac{m^2 (1+r)^2\Gamma}{
2\alpha\beta A
\left(
1+r + \frac{N_\text{eff}/a_\text{eff}}{n_g/a_g}
\right)
}
\left(\frac{a_g}{a_\text{eff}}\right)^2
\frac{M_g^2}{\rho_\chi + P_\chi}
& (\text{Branch II})
\end{cases}
\end{equation}
For the branch I, $m_\text{eff}$ and $c_V^2$ converges to constants for generic choice of the graviton mass parameters $\alpha_{i=0,1,2,3,4}$, and also gradient instability is absent if $\Gamma>0$ in addition to the no-ghost conditions is satisfied.
The same statement applies also to the branch II, that is, the gradient instability is avoided if $\Gamma>0$. 

For the branch II, the sound speed generically becomes 
divergent
at late times as $c_V^2 \propto (\rho_\chi + P_\chi)^{-1}$, indicating the presence of strong coupling 
and superluminality
in this model. This is reminiscent of the vanishing kinetic terms in the minimally coupled case; the late time limit assumed in (\ref{lowElimit_II}) effectively reintroduces the very problems we wish to avoid. 
Such a divergence can be avoided if $\Gamma$ converges to zero at late time, which may be realized if the parameters $\alpha_i$ are fine-tuned. A drawback of such a fine-tuning is that the effective graviton mass $m_\text{eff}^2 \simeq m^2 \Gamma$ becomes zero and then the effect of the graviton mass would be negligible at late time.
Alternatively, further tuning of the evolution such that $r/A^3\gg 1$ would result in a large mass gap larger than the strong coupling scale in the vector sector, as we can see from Eq.~(\ref{calC}). Finally, one can also choose the parameters such that $\rho_\chi+P_\chi$ never becomes too small. In other words, it is possible to discard the assumption that a regime (\ref{lowElimit_II}) exists. 
For example, such a situation is realized if we choose
the parameters such that $\alpha_1+3\alpha_2+3\alpha_3+\alpha_4>0$, $\alpha_2+2 \alpha_3+\alpha_4<0$ and $\alpha_3+\alpha_4>0$. As a result $J(A)>0$ for any value of $A$ and through the constraint (\ref{eq:branch2}), $P_\chi$ is bounded from below parametrically at order $m_g^2M_g^2$. As a result, the $\chi$ matter never becomes subdominant over the two--metric interaction term. 
In the massive gravity version of this theory, the cosmological solutions studied in Ref.~\cite{Gumrukcuoglu:2014xba} have a similar behavior. Both $\rho_\chi$ and $P_\chi$ are constrained to be of order $m^2M_g^2$ and generically there is no $\chi$--field domination nor subdomination over the mass term.

In the early time regime for the branch I, the no-ghost condition is kept satisfied as long as $J>0$ is maintained. However, $m_\text{eff}^2$ and then $c_V^2$ become negative when $P_\chi$ is sufficiently large, as we can see in equation~(\ref{meff}) in which $\Gamma$ stays finite since $A$ converges to a constant in the early time regime. Thus there would be a gradient instability in the early time regime for the branch I.

\subsection{Scalar perturbations}
As next, we will concentrate on the stability of the scalar perturbations. The metric perturbations can be decomposed into scalar perturbations as
\begin{eqnarray}
V_{i,\vec{k}}=ik_i B_{\vec{k}}\,,\qquad
h_{ij,\vec{k}}=2 \delta_{ij} \psi_{\vec{k}}-\left(k_i k_j-\frac{\delta_{ij}}{3} k^2\right)E_{\vec{k}}\,,\nonumber\\
v_{i,\vec{k}}= ik_i b_{\vec{k}}\,,\qquad\;\;
\gamma_{ij,\vec{k}}=2 \delta_{ij} \Sigma_{\vec{k}}-\left(k_i k_j-\frac{\delta_{ij}}{3} k^2\right)S_{\vec{k}}\,.
\end{eqnarray}
As one can see nine degrees of freedom appear in form of a scalar field $\Phi, \varphi, B, b, E, S, \Sigma, \psi$ and $\delta\chi$ from which only two are dynamical. In order to see this we will integrate out the non-dynamical degrees of freedom and also fix the gauge. First of all, let us introduce the following gauge invariant variables ~\cite{DeFelice:2014nja}
\begin{align}
Y_{1,\vec{k}} &= \delta \chi_{\vec{k}} - \frac{\dot\chi_0}{N_\text{eff} H_g}
\left\{
\frac{k^2}{6}\left( \alpha E_{\vec{k}} + \beta A S_{\vec{k}} \right)
+ \left(
\alpha \psi_{\vec{k}}  + \beta A \Sigma_{\vec{k}} 
\right)
\right\}
\notag\\
Y_{2,\vec{k}} &=
S_{\vec{k}} -E_{\vec{k}} 
\notag \\
Y_{3,\vec{k}} &=
\Sigma_{\vec{k}}  - \psi_{\vec{k}}  - \frac{r-1}{N_\text{eff}}
\left\{
\frac{k^2}{6}\left( \alpha E_{\vec{k}}  + \beta A S_{\vec{k}}  \right)
+ \left(
\alpha \psi_{\vec{k}}  + \beta A \Sigma_{\vec{k}} 
\right)
\right\}
\end{align}
We have the freedom to choose a gauge and remove two degrees of freedom. We choose the gauge fixing in a $\alpha$--$\beta$ symmetric way as follows
\begin{equation}
E_{\vec{k}} = - \frac{\beta A}{\alpha+\beta A} \tilde E_{\vec{k}}~,
\quad
S_{\vec{k}} = \frac{\alpha}{\alpha+\beta A} \tilde E_{\vec{k}}~,
\quad
\psi_{\vec{k}} = - \frac{\beta A}{\alpha+\beta A} \tilde \psi_{\vec{k}}~,
\quad
\Sigma_{\vec{k}} = \frac{\alpha}{\alpha+\beta A} \tilde \psi_{\vec{k}}~.
\end{equation}
Under this gauge condition, the dynamical variables reduce to $(Y_1,Y_2,Y_3)=(\delta\chi,\tilde E, \tilde\psi)$. We can further integrate out the non-dynamical degrees of freedom $\Phi$, $\phi$, $B$ and $b$. After using their equations of motion, we find that $Y_3$ is missing the kinetic term and is non-dynamical. It is identified as the degree of freedom of the would-be BD ghost. The cumbersome mathematical expressions involved make it difficult to present it in a manageable way. In the following, we will concentrate on the UV limit and for simplicity set $\Lambda_g= \Lambda_f =0$ without loss of generality. The kinetic term in the UV limit after integrating out the would-be BD mode becomes
\begin{equation}
\mathcal{S}_{\rm scalar}^{(2)}\ni \frac{M_g^2}{2}\int dt d^3 k n_g\,a_g^3
\left(
  \frac{\dot Y_{\vec{k}}}{n_g} {\cal K} \frac{\dot Y_{\vec{k}}^\star}{n_g}
\right)
~,
\end{equation}
where the components of the kinetic matrix are given by
\begin{align}
 {\cal K}_{11} &=
\frac{1}{c_\chi^2 X_\chi}
\frac{a_\text{eff}^3}{a_g^3}
\frac{n_g}{N_\text{eff}}
\frac{\rho_\chi + P_\chi}{M_g^2}
+ {\cal O}(k^{-2})
\nonumber
\\
 {\cal K}_{12} &= {\cal O}(k^{0})\nonumber
\\
 {\cal K}_{22} &=
\frac{k^2 \alpha^2\beta^2A^2 c_V^2 a_\text{eff}^4 (P_\chi + \rho_\chi)^2}{
2M_g^2(r + 1)
} \left[
4 \alpha \beta A c_V^2 a_\text{eff}^2(P_\chi + \rho_\chi) 
- 
M_g^2 a_g^2m_\text{eff}^2(r+1)
\right]^{-1}
+ {\cal O}(k^{0})~.
\label{eq:noghostscalar}
\end{align}
In the UV regime, we find that the kinetic matrix is diagonal at leading order. Therefore, the conditions for avoiding the ghost instability are given by ${\cal K}_{11}>0$ and ${\cal K}_{22}>0$. The latter one is equivalent to
\begin{equation}
4 \alpha \beta A c_V^2 a_\text{eff}^2(P_\chi + \rho_\chi) 
- 
M_g^2 a_g^2m_\text{eff}^2(r+1)
>0~,
\end{equation}
if the stability condition for the vector perturbations $c_V^2>0$ is imposed.

Let us examine the no-ghost condition in each solution branch.
In both branches, the condition ${\cal K}_{11}>0$ is satisfied if the energy condition $\rho_\chi + P_\chi > 0$ is satisfied. As for the other condition ${\cal K}_{22}>0$, in the late time regime for the branch I it reduces to
\begin{equation}
{\cal K}_{22} \simeq 
- 
\frac{k^2(P_\chi + \rho_\chi)^2}{2M_g^4}
\frac{ \alpha^2\beta^2A  a_\text{eff}^4 }{
(r + 1)^2 a_g^2 m^2J
}
<0~,
\end{equation}
which indicates that this branch is plagued by the ghost instability in the late time regime.
On the other hand, in the early time regime where $\rho_\chi$ and $P_\chi$ are much larger than $m^2M_g^2$, the kinetic term ${\cal K}_{22}$ reduces to
\begin{align}
{\cal K}_{22} &\simeq
\frac{k^4\alpha\beta A a_\text{eff}^2 (\rho_\chi + P_\chi)^2}{4M_g^4} 
\left[
\left(1 + r - \frac{N_\text{eff}/ a_\text{eff}}{n_g / a_g}\right)
(\rho_\chi + P_\chi)
+ (1+r)P_\chi
\right]^{-1}
\notag \\
&\simeq
\frac{k^4\alpha\beta A a_\text{eff}^2 (\rho_\chi + P_\chi)^2}{4M_g^4} 
\left[
\frac{r + M_f^2 A^2 / M_g^2}{1 + M_f^2 A^2 / M_g^2} (\rho_\chi + P_\chi)
+ ( 1+r) P_\chi
\right]^{-1}>0~,
\end{align}
where we used $\alpha \simeq \frac{\beta M_g^2}{M_f^2A}$ that holds in this regime to show the latter equality.
Thus the branch-I solutions are free from the ghost instability in the early time regime.

For the branch-II solutions, using equation (\ref{eq:branch2}) we can show that ${\cal K}_{22}$ reduces to
\begin{equation}
{\cal K}_{22} =
\frac{k^2(P_\chi + \rho_\chi) }{4 M_g^2}
\frac{
\alpha \beta A  a_\text{eff}^3
}{
 a_g (\alpha r + \beta A)
}>0~,
\end{equation}
and thus the branch II is free from the ghost instability if the energy condition $P_\chi + \rho_\chi>0$ is satisfied. 
Although we have only obtained the large momentum expressions for the kinetic terms, one can expect that in principle, this branch can lead to a strong coupling at late times.
However, as discussed in the previous Subsection, one can either generate a mass gap larger than the strong coupling scale, or prevent the $\chi$ field to ever become subdominant against the two--metric interaction term, by carefully choosing the parameters of the theory.

\section{Conclusions}
\label{sec:conclusion}

We have studied the cosmological perturbations in the massive bimetric theory with the new matter coupling proposed in \cite{deRham:2014naa}. The background equations give rise to a constraint (\ref{constraint}) which divides the solutions into two branches. We summarized the late time behaviors of the two solution branches, while the early time behaviors were given with the cautious assumption that the doubly coupled matter $\chi$ is the only matter field in the model.
The branch-I solutions defined by $H_g - A H_f=0$ has been studied by \cite{Enander:2014xga}. The second branch defined by $m^2 J - \frac{\alpha\beta a_\text{eff}^2}{M_g^2a_g^2}P_\chi = 0$ is typically dismissed 
on the basis that various problems including strong coupling and nonlinear instability may be present. 
However, we find that this branch may give a well-behaved cosmology at late times; at early times however, as the doubly coupled matter is not allowed to dominate, the standard model fields need to be coupled minimally to one of the metrics.

In the latter part of this work, we conducted quadratic cosmological perturbations and clarified the conditions to avoid the ghost and gradient instabilities for a general background, and then analyzed those conditions for each solution branch focusing on the early and late time regime.
We find that the tensor perturbations are free from both ghost and gradient instabilities if the energy conditions ($\rho_\chi>0$, $\rho_\chi + P_\chi>0$) and some additional conditions ($\Gamma>0$, $J>0$) are satisfied, while the vector sector for the branch I may suffer from the gradient instability in the early time regime.
About the scalar sector, we analyzed the no-ghost conditions and found that the energy conditions guarantee the positivity of one of the kinetic terms, while the second scalar mode becomes a ghost in the late time regime of the branch-I solutions.

In the present work we have assumed a single matter field which coupled to both metric simultaneously. In the presence of additional matter minimally coupled to a single metric, the background solutions at early times would be modified. Generalization to such a case with additional matter fields would be an interesting target for further studies. The late time regime in the present study however provides a good approximation for some of the degrees of freedom of such a complete model, since the background evolution is dominantly governed by the mass term and the effect of the additional matter fields would be small.
Also, due to complexity of the non-diagonal and time dependent action, we were not able to fully diagonalize the scalar sector. Therefore, we cannot present the frequency eigenstates nor address the question of gradient stability for these modes. It would be important to clarify these conditions to examine viability of the branch-II solution which is free from the ghost instability.

\vspace{.5cm}

{\bf Note added:} While this work was being completed, we became aware of Ref.~\cite{Comelli:2015pua}, which also studied the perturbative stability of the bimetric theory with doubly coupled matter. The perturbation analysis in Ref.~\cite{Comelli:2015pua} is performed at the level of the equation of motion, while the present study has been performed directly at the level of the action. 
The results are in 
agreement, while ghost instability in the late time regime of the branch-I solutions is shown for the first time in this work.

\acknowledgments 

AEG acknowledges financial support from the European Research
Council under the European Union's Seventh Framework Programme
(FP7/2007-2013) / ERC Grant Agreement n. 306425 ``Challenging General
Relativity''.  
SM was supported in part by Grant-in-Aid for Scientific Research
24540256 and WPI Initiative, MEXT Japan. Part of his work has been done 
within the Labex ILP (reference ANR-10-LABX-63) part of the Idex SUPER,
and received financial state aid managed by the Agence Nationale de la
Recherche, as part of the programme Investissements d'avenir under the
reference ANR-11-IDEX-0004-02. He is thankful to Institut Astrophysique
de Paris for warm hospitality. 
N.T.\ was supported in part by the European Research Council grant no.\ ERC-2011-StG 279363-HiDGR.

\appendix

	\bibliographystyle{JHEPmodplain}
	\bibliography{EffectiveCoupling_bimetric_cosmology}

\end{document}